\documentclass{article}

\usepackage{times}
\usepackage{bm}
\usepackage[round]{natbib}

\usepackage{amsmath,amssymb}
\usepackage{graphics,graphicx}
\usepackage{mathtools}
\usepackage[tone]{tipa}
\usepackage{bm}
\usepackage{kashsty}
\usepackage{booktabs}
\usepackage{caption}
\usepackage{url}
\usepackage{authblk}

\def\BIBDIR{.}

\usepackage[margin=0.9in]{geometry}

\def\PICDIR{.}

\def\phnaa{\texttt{/\textipa{A}/} }
\def\phnao{\texttt{/\textipa{O}/} }
\def\phndc{\texttt{/\textipa{d}/} }
\def\phniy{\texttt{/\textipa{i}/} }
\def\phnsh{\texttt{/\textipa{S}/} }

\def\phnaax{\texttt{/\textipa{A}/}}
\def\phnaox{\texttt{/\textipa{O}/}}
\def\phndcx{\texttt{/\textipa{d}/}}
\def\phniyx{\texttt{/\textipa{i}/}}
\def\phnshx{\texttt{/\textipa{S}/}}

\def\prob#1{\mathrm{pr}\left(#1\right)}
\def\Prob#1#2{\mathrm{pr}(#1\mid#2)}
\def\xv#1{\mathrm{E}\left(#1\right)}
\def\tr#1{\mathrm{tr}(#1)}

\title{ 
  Inference on covariance operators 
  via concentration inequalities: 
  k-sample tests, classification, and
  clustering via Rademacher complexities
}
\author[1]{Adam B. Kashlak}
\author[2]{ John A. D. Aston }
\author[2]{Richard Nickl}
\affil[1]{\small
  Cambridge Centre for Analysis, 
  University of Cambridge, 
  Wilberforce Rd, Cambridge, CB3~0WB, U.K
}

\affil[2]{\small
  Statistical Laboratory,
  University of Cambridge, 
  Wilberforce Rd, Cambridge, CB3~0WB, U.K
}

\begin{document}

\maketitle

\begin{abstract}
  We propose a novel approach to the analysis of covariance
  operators making use of concentration inequalities.  First,
  non-asymptotic confidence sets are constructed for such
  operators.  Then, subsequent applications including 
  a $k$ sample test for equality of covariance, a
  functional data classifier, and an expectation-maximization style
  clustering algorithm
  are derived and tested on both simulated and phoneme data.
\end{abstract}

\section{Introduction}

Functional data spans many realms of applications from 
medical imaging, \citep{JIANG2014}, 
to speech and linguistics, \citep{PIGOLI2014}, 
to genetics, \citep{PANARETOS2010}.
General inference techniques for functional data are
one area of analysis that has received much attention 
in recent years from the construction of confidence sets,
to other topics such as $k$-sample tests, classification, and 
clustering of functional data.  Most testing methodology treats the
data as continuous $L^2$ valued functions and subsequently 
reduces the problem to a finite dimensional one through 
expansion in some orthogonal basis such as the 
often utilized
\karloeve expansion \citep{HORVATHKOKOSZKA2012}.  
However, inference in the more general Banach space
setting, making use of a variety of norms,
has not been addressed adequately.
We propose a novel methodology for performing fully functional
inference through the
application of concentration inequalities; for general 
concentration of measure results, see
\cite{LEDOUX2001} and \cite{BOUCHERON2013}. 
Special emphasis is given to inference on covariance
operators, which offers a fruitful way to analyze 
functional data.

Imagine multiple samples of speech data collected from
multiple speakers.  Each speaker will have his or her own 
sample covariance operator taking into account the unique
variations of his or her speech and language.  
An exploratory researcher may want to find natural clusters
amidst the speakers perhaps corresponding to gender, 
language, or regional accent.  Meanwhile, a linguist studying
the similarities between languages may want to test
for the equality of such covariances.
A computer scientist may need to implement an algorithm
that when given speech data quickly identifies what 
language is being spoken and furthermore parses the sound clip and
identifies each
individual phoneme in order to process the speech into text.
Our proposed method has the versatility to yield statistical tests that
answer these questions as well as others.

Past methods for analyzing covariance operators 
\citep{PANARETOS2010,FREMDT2013} 
rely on the Hilbert-Schmidt 
setting for their inference.  However, the recent
work of \citet{PIGOLI2014} argues that 
the use of the Hilbert-Schmidt metric ignores the 
geometry of the covariance operators and that more
statistical power can be gained by using alternative 
metrics.
Hence, we approach such inference for covariance operators
in the full generality of Banach spaces, 
by using a non-asymptotic concentration of measure approach, 
which has previously been used in nonparametric statistics 
and machine learning, sometimes under the name of 
`Rademacher complexities' 
\citep{
  KOLTCHINSKII2001,KOLTCHINSKII2006,BARTLETT2002,
  BARTLETT2003,GINENICKL2010CONFIDENCE,
  ARLOT2010,LOUNICINICKL2011,KERKYACHARIAN2012,FANPARTIII}.
These concentration inequalities
provide a natural way to construct non-asymptotic confidence
regions.

\section{Definitions and notation}
\label{sec:defns}


Generally, we will consider functional data to 
be in the Hilbert space $L^2(I)$ for $I\subset\real$.
Our estimated covariance operators of interest are 
treated as Banach space valued random variables. 
Let 
$$
  Op(L^2) = \left\{
    T:L^2(I)\rightarrow L^2(I)\,\middle|\,
    \exists M\ge0, \norm{T\phi}_{L^2} \le M\norm{\phi}_{L^2}~
    \forall \phi \in L^2(I)
  \right\}
$$ 
denote the space of all bounded linear operators 
mapping $L^2$ into $L^2$.  This is where our covariance
operators will live.  


The metrics that will be investigated are those 
that correspond to the {p-Schatten norms}.  When
$p\neq2$, these are not Hilbert norms.  Hence,
we will consider the general Banach space setting.

\begin{definition}[$p$-Schatten Norm]
  Given separable Hilbert spaces $H_1$ and $H_2$, 
  a bounded linear operator $\Sigma:H_1\rightarrow H_2$,
  and $p\in[1,\infty)$,
  then the $p$-Schatten norm is 
  $
    \norm{\Sigma}^p_{p} = \tr{(\Sigma^*\Sigma)^{p/2}}.
  $
  For $p=\infty$, the Schatten norm is the operator norm:
  $
    \norm{\Sigma}_\infty = \sup_{f\in H_1}( 
    {\norm{\Sigma f}_{H_2}}/{\norm{f}_{H_1}}).
  $
  In the case that $\Sigma$ is compact, self-adjoint, and trace-class,
  then given the associated eigenvalues $\{\lmb_i\}_{i=1}^\infty$,
  the $p$-Schatten norm coincides with the $\ell^p$ norm of the 
  eigenvalues:
  $$
    \norm{\Sigma}_{p}^p =\left\{ 
    \begin{array}{ll}
      \norm{\lmb}_{\ell^p}^p = \sum_{i=1}^\infty \abs{\lmb_i}^p, & 
      p\in[1,\infty) \\
      \max_{i\in\natural} \abs{\lmb_i}, & p=\infty
    \end{array}
    \right.
  $$
\end{definition}

In order to construct a covariance operator from 
a sample of functional data, the notion of 
{tensor product} 
is required.
Let $f,g\in L^2(I)$ 
and $\phi$ in the dual space $L^2(I)^*$
with inner product
$\iprod{f}{\phi} = \phi(f)$. 
The tensor product,
$f\otimes g$, is the rank one operator defined by
$(f\otimes g)\phi = \iprod{g}{\phi}f = \phi(g) f$.

Secondly, we will implement a Rademacher symmetrization 
technique in the concentration inequalities to come.  This
requires the use of the namesake Rademacher random variables.
\begin{definition}[Rademacher Distribution]
  A random variable $\veps\in\real$ has a 
  {Rademacher} distribution
  if 
  $
    \prob{\veps=1} = \prob{\veps=-1} = 1/2.
  $
\end{definition}


One particularly fruitful avenue of functional data analysis 
is the analysis of {covariance operators}.
Such an approach to functional data has been discussed 
by \cite{PANARETOS2010} for DNA microcircles,
by \cite{FREMDT2013} for the egg laying practices 
of fruit flies, and by \cite{PIGOLI2014} 
with application to differentiating spoken languages.  

\begin{definition}[Covariance Operator]
  Let $I \subseteq \real$, and let $f$ be a random function 
  (variable) in $L^2(I)$ with $\xv{\norm{f}_{L^2}^2}<\infty$ and mean zero.
  The associated 
  {covariance operator} $\Sigma_f\in Op(L^2)$ is defined 
  as
  $
    \Sigma_f = \xv{ \tensor{f}{2}} = \xval{ \iprod{f}{\cdot}f }.
  $
\end{definition}

If $I = \{i_1,\ldots,i_m\}$ has finite cardinality, then
$f=(f_1,\ldots,f_m)$ is a random vector in $\real^m$ 
and for some fixed vector $v\in\real^m$, 
$\xv{\iprod{f}{v}f} = \xv{f\TT{f}}v$ where 
$\Sigma_f=\xval{f\TT{f}}$ is the usual covariance matrix.
Covariance operators are integral operators with the 
kernel function $c_f(s,t) = \mathrm{cov}\{{f(s)},{f(t)}\}\in{L^2(I\times I)}$.
Such operators are characterized by
the result that for $f\in L^2(I)$, $\Sigma_f$ is a covariance 
operator \iff it is trace-class, self-adjoint, and compact on $L^2(I)$
where the symmetry follows immediately from the definition and
the finite trace norm comes from Parseval's equality
\citep[Theorem 1.7]{BOSQ2012}\citep[Section 2.3]{HORVATHKOKOSZKA2012}.

Furthermore, working under the assumption that 
$\xv{\norm{f}_{L^2}^4}<\infty$, 
we will require tensor powers of covariance
operators denoted as
$\tensor{\Sigma}{2}:Op(L^2)\rightarrow Op(L^2)$.  For
a basis $\{e_i\}_{i=1}^\infty\in L^2(I)$ with corresponding 
basis $\{e_i\otimes e_j\}_{i,j=1}^\infty$ for $Op(L^2(I))$, the 
previous definition
is extended to $\tensor{\Sigma}{2}=\iprod{\Sigma}{\cdot}\Sigma$
where for $\Sigma_1,\Sigma_2\in Op(L^2)$ with 
$\Sigma_1=\sum_{i,j=1}^\infty \lmb_{i,j}e_{i,j}$ and 
$\Sigma_2=\sum_{i,j=1}^\infty \gamma_{i,j}e_{i,j}$, then
$\iprod{\Sigma_1}{\Sigma_2}=\sum_{i,j}^\infty\lmb_{i,j}\gamma_{i,j}$.  
Specifically for covariance operators, the tensor power takes on 
a similar integral operator form with kernel
$
   c_{\Sigma}(s,t,u,v) = \cov{f(s)}{f(t)}\cov{f(u)}{f(v)}.
$

  Given an Hilbert space $H$ with inner product 
  $\iprod{\cdot}{\cdot}$, the {adjoint} of a bounded linear operator 
  $\Sigma:H\rightarrow H$, denoted as
  $\Sigma^*$, is the unique operator such that
  $
    \iprod{\Sigma f}{g} = \iprod{f}{\Sigma^* g}
  $
  for $f,g\in H$.  The existence of which is given by the
  Riesz representation theorem \citep[chapter 2]{RUDIN1987}.
  For {self-adjoint} operators, such
  as the covariance operators of interest, $\Sigma=\Sigma^*$.
  

We begin with a sample of functional data.  Let 
$f_1,\ldots,f_n\in L^2(I)$ be
\iid observations 
with mean zero
and covariance operator $\Sigma$.  Let the sample mean
be $\bar{f} = {n^{-1}}\sum_{i=1}^n f_i$ and the 
empirical estimate of $\Sigma$ be
$
  \hat{\Sigma} = n^{-1}
  \sum_{i=1}^n ( f_i-\bar{f} )\otimes( f_i-\bar{f} ).
$
The initial goal is to construct a confidence set for ${\Sigma}$
with respect to some metric $\mathrm{d}({\cdot},{\cdot})$
of the form
$
  \{ {\Sigma} : \mathrm{d}({\hat{\Sigma}},{\Sigma}) 
  \le r(n,\hat{\Sigma},\alpha) \},
$
which has coverage $1-\alpha$ for any desired $\alpha\in[0,1]$ and 
a radius $r$ depending only on the data and $\alpha$.
Such a confidence set can be utilized for a wide variety 
of statistical analyses.

\section{Constructing a confidence set}
\label{sec:confSet}

\subsection{Confidence sets for the mean in Banach spaces}

The goal of this section is to construct a non-asymptotic 
confidence region in the Banach space setting.
These can subsequently be specialized
to our case of interest, covariance operators, when 
the $X_i$ below are replaced with $\tensor{f_i}{2}$.

The construction of our confidence set is based on 
Talagrand's concentration inequality \citep{TALAGRAND1996CON} with 
explicit constants. This inequality is typically stated 
for {empirical processes} 
\citep[Theorem 3.3.9 and 3.3.10 ]{GINENICKL2015}, but applies to random 
variables with values in a separable Banach space $(B, \norm{\cdot}_B)$ 
as well by simple duality arguments 
\citep[Example 2.1.6]{GINENICKL2015}.
Let $X_1,\ldots,X_{n}\in(B,\norm{\cdot}_B)$ be mean zero \iid 
Banach space valued random variables with $\norm{X_i}_B\le U$ for all
$i=1,\ldots,n$ where $U$ is some positive constant.  Furthermore, let 
$\iprod{\cdot}{\cdot}:B\times B^* \rightarrow \real$ such that
for $X\in B$ and $\phi\in B^*$ then $\iprod{X}{\phi}=\phi(X)$.  
Define
\begin{align*}
  Z &= \sup_{\norm{\phi}_{B^*}\le1} \sum_{i=1}^{n} \iprod{X_i}{\phi}
       = \norm{\sum_{i=1}^{n} X_i}_B, &
  \sigma^2 &= 
       \frac{1}{n}\sum_{i=1}^{n}\sup_{\norm{\phi}_{B^*}\le1}
       \xv{\iprod{X_i}{\phi}^2},
\end{align*}
where the supremum is taken over a countably dense subset of 
the unit ball of $B^*$.
Furthermore, define $v = 2U\xv{ Z} + n\sigma^2$.  Then, 
$
  \mathrm{pr}\{{Z}>\xv{Z}+r\} \le
  \exp\{{-r^2}/({2v + 2rU/3})\}.
$
Rewriting $Z$ as $n\norm{\bar{X}-\xv{ \bar{X}}}_{B}$ results in
$$
  \mathrm{pr}\left[
    \norm{\bar{X}-\xv{ \bar{X}}}_{B}>\mathrm{E}\left\{
      \norm{\bar{X}-\xv{\bar{X}}}_{B}
    \right\}+r
  \right] <
  \exp\left(\frac{-n^2r^2}{2v + 2nrU/3}\right)
$$
where $\norm{X_i}_{B}<U$ and 
$v = 2nU\mathrm{E}\{\norm{\bar{X}-\xv{\bar{X}}}_{B}\} + n\sigma^2$.

The above tail bound incorporates the unknown 
$\mathrm{E}(\lVert{\bar{X}-\xv{\bar{X}}}\rVert_{B})$.  Consequently, 
a {symmetrization} technique is used.
This term is 
replaced by the norm of the Rademacher average 
$
  R_n = {n^{-1}}\sum_{i=1}^n\veps_i(X_i- \bar{X})
$
where the $\veps_i$ are \iid Rademacher random variables 
also independent of the $X_i$.
This substitution is justified by 
invoking the symmetrization inequality \citep[Theorem~3.1.21]{GINENICKL2015},
$$
  \xv{Z} = 
  \mathrm{E}\left\{
    \norm{ \frac{1}{n}\sum_{i=1}^n(X_i- \xv{ \bar{X}})}_{B}
  \right\} \le
  2\mathrm{E}\left\{
    \norm{ \frac{1}{n}\sum_{i=1}^n\veps_i(X_i- \bar{X}) }_{B}
  \right\} 
  = 2\xv{\norm{R_n}_{B}}.
$$

In practice, the coefficient of $2$ is unnecessary as
averaging the data already has a symmetrizing effect,
and if starting from symmetric data, then 
$X-\bar{X}$ and $\veps(X-\bar{X})$ are equal in distribution.
This result allows us to replace the original expectation with 
the expectation of the Rademacher average.
Furthermore, Talagrand's inequality also applies to $R_n$.
Hence, the Rademacher average concentrates strongly about its
expectation, which justifies dropping the expectation.
In practice, one can use the 
intermediary $\xvsub{\veps}{\norm{R_n}_{B}}$, which can be approximated  
for reasonable sized data sets via
Monte Carlo simulations of the $\veps_i$.  However, this 
is not strictly necessary, and for large data sets, a single
random draw of $\veps_i$ will suffice \cite[Section~3.4.2]{GINENICKL2015}.  

The resulting $(1-\alpha)$-confidence set is
\begin{equation}
  \label{eqn:confSet}
  \left\{ {X} :
    \norm{{X}-\bar{X}}_{B} \le
    \norm{R_n}_{B} +
    \left\{\frac{2}{n}\log(2\alpha)\left(
      \sigma^2 + 2U\norm{R_n}_{B}
    \right)\right\}^{1/2} +
    \frac{U\log(2\alpha)}{3n}
  \right\}.
\end{equation}
To make use of these results in practice, both the weak variance
$\sigma^2$ must be estimated for the data and a reasonable
choice of $U$ must be made, and a main contribution of this 
present paper is to propose some theoretically motivated but 
practically useful non-asymptotic choices 
for these constants that work for the functional data applications 
we are investigating.  This will be discussed in the
setting of covariance operators in the following subsection.

\subsection{ Confidence sets for covariance operators }
\label{sec:confSetCov}

To construct a confidence set for covariance operators, 
let our functional data $f_i\in L^2(I)$ and
$\tensor{f_i}{2} = f_i\otimes f_i\in Op(L^2)$, 
the Hilbert space of bounded linear operators 
mapping $L^2$ to
$L^2$, such that $(f_i\otimes f_i)\phi = \iprod{f}{\phi}f$ for some 
$\phi\in L^2$.
Hence, for some desired $p$-Schatten norm, $\norm{\cdot}_p$, 
with $p\in[1,\infty)$ and with conjugate $q=p/(p-1)$, 
we have
\begin{align*}
  Z &= \norm{\frac{1}{n}\sum_{i=1}^n f_i\otimes f_i - \xv{f_i\otimes f_i}}_p 
    = \norm{\hat{\Sigma}-\Sigma}_p,
  &
  \sigma^2 
  &= \frac{1}{n}\sum_{i=1}^n\sup_{\norm{\Pi}_q\le1}
     \mathrm{E}\left\{
       \iprod{f_i^{\otimes2}-\xv{ f_i^{\otimes2}}}{\Pi}^2
     \right\}
\end{align*}
for the supremum being taken over a countably dense subset of
the unit ball of $\Pi\in Op(L^2)$. 
The Rademacher average,
$
  R_n = {
    {n^{-1}}\sum_{i=1}^n \veps_i\{
      \tensor{(f_i-\bar{f})}{2}-
      \hat{\Sigma}
    \}
  },
$
is also in $Op(L^2)$, because for any $\phi\in L^2(I)$ and 
for some $M\in\real$,
$
  \norm{R_n \phi}_{L^2} \le \abs{\veps_i}\lVert{ 
    \{\tensor{(f_i-\bar{f})}{2}-\hat{\Sigma} \}\phi
  }\rVert_{L^2} \le M\norm{\phi}_{L^2}
$
since $\{\tensor{(f_i-\bar{f})}{2}-\hat{\Sigma} \}$ is
a bounded operator and $\abs{\veps_i}=1$.
Furthermore, in the case that there exists a fixed $c\in\real$ with 
$\norm{f_i}_{L^2}\le c$ for all $i$ corresponding to a physical 
bound on the energy of $f_i$, 
$\lVert{\tensor{f_i}{2}}\rVert_p = \lVert{f_i}\rVert_{L^2}^2 \le c^2 = U$.
It will be determined below that $U\ge\sigma$ in this case.  
In general, setting $U=\sigma$  
gives good experimental results when $f_i$ is Gaussian
as will be discussed in later sections.
This
results in $v_n \approx \sigma^2/n$.
For any $p\in[1,\infty)$ and $\alpha\in[0,1/2]$,
the proposed $(1-\alpha)$-confidence set inspired by 
Equation~\ref{eqn:confSet} for covariance operators is
\begin{equation*}
  C_{n,1-\alpha} = \left[ {\Sigma} :
    \norm{\hat{\Sigma}-{\Sigma}}_p \le
    \norm{R_n}_p +
    \sigma\left\{-\frac{2}{n}\log(2\alpha)\right\}^{1/2}-
    \frac{\sigma\log(2\alpha)}{3n}
  \right].
\end{equation*}
where $\sigma$ depends on the distribution on the functional data.
As a rule of thumb for the choice of $\sigma^2$, as we will now show,
is to note that 
$\sigma^2 \le \lVert{ \mathrm{E}( f^{\otimes 4}) - \Sigma^{\otimes2} }\rVert_p$
and to estimate this bound empirically by $\hat{\sigma}^2$.
For example, when the $f_i$ are from a Gaussian process
$\hat{\sigma} \le {2^{1/2}}\lVert{\Sigma}\rVert_p$ as explained in detail
in Section~\ref{sec:sigmaGauss}.  

To calculate the weak variance $\sigma^2$, define 
$f^{\otimes n}=f\otimes\ldots\otimes f$ to be the n-fold tensor
product of $f$ with itself and extend the definition of 
$
 \iprod{\cdot}{\cdot}:
 \tensor{(L^2)}{4}\times\{\tensor{(L^2)}{4}\}^*\rightarrow \real
$
such that 
$
  \iprod{\tensor{f}{4}}{\tensor{\phi}{4}} = 
  \iprod{\tensor{f}{2}}{\tensor{\phi}{2}}^2 = 
  \iprod{f}{\phi}^4 = \phi(f)^{4}.
$
For operators $\Pi\in\{\tensor{(L^2)}{2}\}^*$ and 
$\Xi\in\{\tensor{(L^2)}{4}\}^*$,
the weak variance is 
\begin{align*}
  \sigma^2 
  &= \frac{1}{n}\sum_{i=1}^n\sup_{\norm{\Pi}_q\le1}
     \mathrm{E}\left\{
       \iprod{f_i^{\otimes2}-\xv{ f_i^{\otimes2}}}{\Pi}^2
     \right\}\\
  &\le \frac{1}{n}\sum_{i=1}^n\sup_{\norm{\Xi}_q\le1}
     \iprod{
       \xv{ f_i^{\otimes4}} - \left\{\xv{ f^{\otimes2}}\right\}^{\otimes2}
     }{\Xi}
  \le \norm{ \xv{ f^{\otimes 4}} - \Sigma^{\otimes2} }_p
\end{align*}
where the inequality stems from the fact that the supremum is
being taken over a larger set.
However, in the Hilbert space setting, the dual of the tensor product does 
coincide with the tensor product of the dual space, and thus the 
above inequality can be replaced with an equality if the Hilbert-Schmidt
norm, 2-Schatten norm, is used.  Given a bound 
$\norm{f_i}_{L^2}^2\le c^2=U$, then
$
  \sigma^2\le\lVert{ \xv{ f^{\otimes 4}}}\rVert_p\le
  \mathrm{E}(\lVert{f}\rVert_{L^2}^4) \le c^4 = U^2.
$



\subsection{The weak variance for $p=\infty$}

Let $E$ be a countable dense subset of the unit ball of $L^2(I)$.
In the case $p=\infty$, we cannot use duality, but can still write
$Z$ and $\sigma^2$ as suprema over the countable set
and achieve the same results as above.
\begin{align*}
  Z &= 
  \frac{1}{n}\sup_{e\in E} \sum_{i=1}^n 
    \iprod{\left\{\tensor{f}{2}_i-\xv{\tensor{f}{2}_i}\right\}e}{e}
  = \sup_{e\in E}\iprod{(\hat{\Sigma}-\Sigma)e}{e}
  = \norm{\hat{\Sigma}-\Sigma}_\infty, \\
  \sigma^2 &= 
  \frac{1}{n} \sum_{i=1}^n \sup_{e_1\in E}\mathrm{E}\left\{
    \iprod{(\tensor{f_i}{2}-\Sigma)e_1}{e_1}^2
  \right\} \le
  \frac{1}{n} \sum_{i=1}^n \sup_{e_1,e_2\in E}\mathrm{E}\left\{
    \iprod{\tensor{f_i}{2}-\Sigma}{e_1\otimes e_2}^2 
  \right\} \\
  &\le \frac{1}{n} \sum_{i=1}^n \sup_{e_1,e_2\in E}
    \iprod{\left\{
      \xv{\tensor{f_i}{4}}-\tensor{\Sigma}{2}
    \right\}(e_1\otimes e_2)}{e_1\otimes e_2}
  = \norm{\xv{\tensor{f_i}{4}}-\tensor{\Sigma}{2}}_\infty.
\end{align*}
As before, if $\norm{\tensor{f_i}{2}}_\infty=\norm{f_i}_{L^2}^2\le c^2=U$,
then $\sigma^2 \le U^2$.

\subsection{The weak variance for Gaussian data}
\label{sec:sigmaGauss}

Similarly to the bounded case, we estimate 
$\norm{ \xv{ f^{\otimes 4}} - \Sigma^{\otimes2} }_p$ for Gaussian data.
Consider $f$ from a Gaussian process
with mean zero and covariance $\Sigma$
and define $f_s = f(s)$.{
  Strictly speaking these variables are not norm bounded, 
  but similar to concentration results for Gaussian random
  variables in $\real^d$ 
  \citep[Theorem~3.1.9]{GINENICKL2015}, we found below that our methods 
  still work well.
}
The integral kernel can be written as \citep{ISSERLIS1918}
\begin{align*}
  \xval{f_sf_tf_uf_v} &= \xval{f_sf_t}\xval{f_uf_v} + 
  \xval{f_sf_u}\xval{f_tf_v} + \xval{f_sf_v}\xval{f_tf_u}\\
  &= c_f(s,t)c_f(u,v)+c_f(s,u)c_f(t,v)+c_f(s,v)c_f(t,u).
\end{align*}
Hence, we have that $
  \xval{f_sf_tf_uf_v}-\Sigma_{s,t}\Sigma_{u,v} = 
  \Sigma_{s,u}\Sigma_{t,v} + \Sigma_{s,v}\Sigma_{t,u}
$ and that the operator $\xv{ f^{\otimes 4}} - \Sigma^{\otimes2}$,
which can be thought of as an Hilbert-Schmidt operator on the
space $Op(L^2)$,
can be represented by the integral kernel 
$c_f(s,u)c_f(t,v)+c_f(s,v)c_f(t,u)$.  These two terms are merely
relabeled versions of $\tensor{\Sigma}{2}$.
Consequently, using the subadditivity of the norm,
$
  \norm{ \xv{ f^{\otimes 4}} - \Sigma^{\otimes2} }_p \le 
  \norm{\Sigma^{\otimes2}}_p + \norm{\Sigma^{\otimes2}}_p=
  2\norm{\Sigma^{\otimes2}}_p.
$
For example, for the Hilbert-Schmidt norm,
\begin{align*}
  \norm{ \xv{ f^{\otimes 4}} - \Sigma^{\otimes2} }^2_{HS} &= 
  \iiiint\left\{
    c_f(s,u)c_f(t,v)+c_f(s,v)c_f(t,u)
  \right\}^2dsdtdudv \\
  &= 2\norm{\Sigma}^4_{HS} + 
     2\iiiint c_f(s,u)c_f(s,v)c_f(t,v)c_f(t,u) dsdtdudv
  \le 4\norm{\Sigma}^4_{HS}.
\end{align*}
Lemma~5.1 of \cite{HORVATHKOKOSZKA2012} gives 
an explicit form of a covariance operator 
of $\Sigma$ in terms of the eigenfunctions of $\Sigma$
for Gaussian data in the Hilbert-Schmidt setting.

Given $\lmb_i$, the eigenvalues of $\Sigma$,
the spectrum of $\Sigma^{\otimes2}$ is $\{ \lmb_i\lmb_j\}_{i,j=1}^\infty$.
Hence, for any of the $p$-Schatten norms,  
$\lVert{\Sigma\otimes\Sigma}\rVert_p = \lVert{\Sigma}\rVert_p^2$. 
Note that in the above calculations, the weak variance depends on the unknown
$\Sigma$.  In practice, this can be replaced by the 
empirical estimate $\hat{\Sigma}$.

\section{Applications}
\label{sec:apps}
\subsection{$k$ Sample Comparison}
\label{sec:kSamp}

Testing for the equality of means among multiple sets of data 
is a common task in data analysis.  In the functional setting,
there has been recent work on performing such a test on 
covariance operators in order to test 
whether or not $k$ sets of curves have similar variation.
\cite{PANARETOS2010} propose such a method for 
a two sample test on covariance operators 
given data from Gaussian processes. 
Similarly, \cite{FREMDT2013}
propose a non-parametric two sample test on covariance operators.  
Both of these
approaches make use of the \karloeve expansion and, 
hence, the underlying Hilbert space geometry. 
\cite{PIGOLI2014} take a comparative look at a variety of metrics
to rank their statistical power when used in a two 
sample permutation test.

Following from the results of~\cite{PIGOLI2014}, our method
uses the $p$-Schatten norms with the concentration inequality
based confidence sets of the previous section
to compare covariance operators.  In the two sample setting,
we are able to achieve similar statistical power to that 
of the permutation test after proper tuning of the coefficients
in the inequalities.  
Furthermore, the analytic nature of the concentration 
approach leads to a significant reduction in computing time,
which offers an even more significant savings for larger 
values of $k$.

From the confidence set constructed in the previous section,
we can devise a test for comparing the empirical covariance operators
generated from $k$ samples of functional data. Let the
$k$ samples be
$f_1^{(1)},\ldots,f_{n_1}^{(1)},\ldots,f_{1}^{(k)},\ldots,f_{n_k}^{(k)}$
where for each sample $i$ and all elements $j=1,\ldots,n_i$, 
$f_{j}^{(i)}$ has covariance $\Sigma^{(i)}$.  
Our goal is to design a test for the following two hypotheses:
\begin{align*}
  \text{H}_0:& ~~\Sigma^{(1)}=\ldots=\Sigma^{(k)}&
  \text{H}_1:& ~~\exists\, i,j \text{ s.t. } \Sigma^{(i)}\ne\Sigma^{(j)}.
\end{align*}

To achieve this, a pooled estimate of the weak variance is
computed as a weighted average of each sample's individual weak 
variance in similar style to that of a standard t-test.  
Let the total data size be $N = n_1+\ldots+n_k$
and $\sigma_i^2$ be the weak variance for sample $i$, then
the pooled variance is defined as
$
  \sigma_\text{pool}^2 = {N^{-1}}\sum_{i=1}^k n_i\sigma_i^2.
$
Given Gaussian data and the $p$-Schatten norm, for example, 
this reduces to 
$
 \sigma_\text{pool}^2 = {2}{N^{-1}}\sum_{i=1}^k 
 n_i\lVert{\Sigma^{(i)}}\rVert_p^2
$.  In practice, $\sigma_\text{pool}^2$ is estimated
from the data for the following confidence regions
in order to have those regions only depend on the data.

Taking inspiration from the standard analysis of variance
\citep[Chapter 11]{CASELLABERGER2002},
let $\hat{\Sigma}^{(i)}$ be the empirical estimate of the 
covariance operator for the $i$th sample, and let $\hat{\Sigma}$
be the estimate of the covariance operator for the total data set.
Making use of the confidence sets for covariance operators
from Section~\ref{sec:confSetCov} gives
the rejection region
\begin{multline*}
  \mathcal{C} = \left\{ f : 
    \sum_{i=1}^k \norm{\hat{\Sigma}^{(i)}-\hat{\Sigma}}_p > 
    \sum_{i=1}^k \norm{ 
      \sum_{j=1}^{n_i} \veps_{i,j}\left(
        \tensor{{f_{j}^{(i)}}}{2} - \hat{\Sigma}
      \right) 
    }_p + \right. \\ + \left.
    \left(\sum_{i=1}^k \frac{\sigma_\text{pool}^2}{n_i}\right)^{1/2}
      \left( -2\log 2\alpha \right)^{1/2} +
    \left(\sum_{i=1}^k \frac{\sigma_\text{pool}}{n_i}\right)
      \frac{\log 2\alpha}{3}
  \right\},
\end{multline*}
which under the null hypothesis will have size no greater than the
desired $\alpha$.  

The size of the test induced by this rejection region is significantly
less than the target size $\alpha$ due to the use of multiple 
concentration inequalities. Hence, tuning the inequalities
is required to yield a useful test.  Many experiments were run
on simulated data sets generated as samples from a Gaussian process
with randomly generated covariance operators whose
eigenvalues were chosen to decay at a variety of rates. 
In this setting, the 
coefficients of $1-k^{-1/2}$ for the Rademacher term 
and $(k+2)/(k+3)$ for the
deviation term were determined experimentally to improve the
size of the confidence region:
\begin{multline}
  \label{eqn:nsampRegion}
  \mathcal{C} = \left[ f : 
    \sum_{i=1}^k \norm{\hat{\Sigma}^{(i)}-\hat{\Sigma}}_p > 
    \left(1-k^{-1/2}\right)
    \sum_{i=1}^k \norm{ 
      \sum_{j=1}^{n_i} \veps_{i,j}\left(
        \tensor{{f_{j}^{(i)}}}{2} - \hat{\Sigma}
      \right) 
    }_p + \right. \\ + \left. \left(\frac{k+2}{k+3}\right)\left\{
    \left(\sum_{i=1}^k \frac{\sigma_\text{pool}^2}{n_i}\right)^{1/2}
      ( -2\log 2\alpha )^{1/2} +
    \left(\sum_{i=1}^k \frac{\sigma_\text{pool}}{n_i}\right)
    \frac{\log 2\alpha}{3}
  \right\}\right].
\end{multline}

\subsection{ Classification of Operators }
\label{sec:class}

Classification of functional data has been an area of heavy
research over the last two decades. 
\cite{JAMESHASTIE2001} extend linear discriminant
analysis to functional data.  \cite{HALL2001} and 
\cite{GLENDINNING2003} classify with principle components.  
\cite{FERRATYVIEU2003} implement kernel 
estimators.  General linear models for functional data are
discussed by 
\cite{MULLERSTADTMULLER2005}.  
\cite{DELAIGLEHALL2012} analyze the asymptotic 
properties of the centroid based classifier.
Wavelet based classification is detailed by 
\cite{BERLINET2008} and \cite{CHANG2014}.

One application of our method beyond classification of 
functional data is the classification of covariance operators.
In the setting of speech analysis, consider multiple 
speakers and multiple samples of speech from each speaker.
The speech samples can be combined into a single sample
covariance operator for each speaker.  Then, our method
can be employed, for example, to classify
the covariance operators by speaker gender or speaker language.
Evidence that this is a fruitful approach can be found in 
the analysis of \cite{PIGOLI2014} and \cite{PIGOLI2015} where a 
variety of metrics are compared for their efficacy when 
performing inference on covariance operators. 
These articles detail
the discrepancy between
sample covariance operators produced by speakers of different
romance languages.

Given $k$ possible labels and $n$ samples of labeled 
data $(Y_i,f_i)$ with label $Y_i\in\{1,\ldots,k\}$ and observation
$f_i\in L^2(I)$, our goal is to determine the probability
that a newly observed $g\in L^2(I)$ belongs to label $Y=j$.
Given such a $g$, the Bayes classifier chooses the label 
$y = \arg \max_{j} \Prob{Y=j}{g}$ where
$ \Prob{Y=j}{g} = {\Prob{g}{Y=j}\prob{Y=j}}/{\prob{g}}$.

Beginning with a training set of $n$ samples with
$n_j$ samples of label $j$, the sample mean of each category 
is computed: $\bar{f}_j = {n_j^{-1}}\sum_{i:Y_i=j}f_i$.
The probability $\Prob{g}{Y=j}$ above is replaced with
$
  \mathrm{pr}[
    \norm{\bar{f}_j-g}_{L^2} > 
    \mathrm{E}\{\norm{\bar{f}_j-\xv{\bar{f}_j}}_{L^2}\}+r
  ]
$
with the goal of making a decision based on how much more 
$\bar{f}_j$ differs from $g$ than $\bar{f}_j$ differs from
its expectation $\xv{\bar{f}_j}$.
Similar techniques to those is Section~\ref{sec:confSet} as used.
Define the {Rademacher sum}, $R_j$, and the {empirical weak variance}, 
$\hat{\sigma}_j^2$, for label $j$ to be, respectively,
\begin{align*}
  R_{j} &= \frac{1}{n_j}\sum_{i:Y_i=j}\veps_i(f_i- \bar{f}_j),&
  \hat{\sigma}_j^2&=\norm{
    \frac{1}{n_j}\sum_{i:Y_i=j} \tensor{f_i}{2} - 
    \tensor{\bar{f}_j}{2}
  }_p
\end{align*}
where $\veps_i$ are \iid Rademacher random variables.
The tail bound for the above probability is then
\begin{equation}
  \label{eqn:conClass}
  \prob{
    \norm{\bar{f}_j-g}_{L^2}-\norm{R_j}_{L^2}>r
  } <
  \exp\left(
    \frac{-n_jr^2}{4\norm{R_j}_{L^2}U + 2
      \hat{\sigma}_j^2+2rU/3
    }
  \right),
\end{equation}
where $U$ is an upper bound on $\norm{f_i}_{L^2}$.
However, this can be approximated by the 
Gaussian tail 
$
  \exp\left(
    -n_jr^2/2\sigma^2_j
  \right).
$
In the simulations of Section~\ref{sec:simData},
this approximation actually achieves a better correct classification rate 
on both Gaussian and t-distributed data.
This specifically works on t-distributed data as the estimate in
Equation~\ref{eqn:labelProb} below is merely concerned with comparing
the tail bounds rather than their specific values.  Consequently, 
the tail for every category is underestimated in the t case, 
but the ratio remains valid for comparison purposes.

Assuming uniform priors on the labels,
the estimate for the probability expression 
in the Bayes classifier is
\begin{align}
  \label{eqn:labelProb}
  \Prob{Y = j}{g} &\approx 
  \frac{\phi_j({g})}{\sum_{l=1}^k \phi_l({g})},
  &
  \phi_j(g) &= 
  \exp\left\{
    -\frac{n_j}{2}\left(
      \frac{\norm{\bar{f}_j-g}_{L^2}-\norm{R_j}_{L^2}}{ \hat{\sigma}_j }
    \right)^2
  \right\}.
\end{align}
This can be extended to the case where an unlabeled observation is a 
collection of curves $g_1,\ldots,g_m$ by replacing
$\lVert{\bar{f}_j-g}\rVert_{L^2}$ in the above expression with
$\lVert{\bar{\Sigma}_j-\hat{\Sigma}_g}\rVert_{p}$ where $\bar{\Sigma}_j$
is the sample covariance of the $f_i$ with label $j$ and
$\hat{\Sigma}_g$ is the sample covariance of the $g_i$.
The Rademacher and weak variance terms would also be updated
accordingly.

\subsection{ Clustering of Operator Mixtures }
\label{sec:emalg}

Closely related to the problem of classification is the
problem of clustering.  Given a sample of functional data,
we want to assign one of a finite collection of labels to each
curve.  For example, in speech processing, one may want to 
cluster sound clips based the language of the speaker, or,
to be discussed in Section~\ref{sec:emAlgPractice}, one may 
want to separate unlabeled phoneme curves into clusters.

There have been many recently proposed methods for clustering
functional data.  Many approaches begin by constructing a low
dimensional representation of the
data in some basis such as 
modelling the data with a B-spline basis 
followed by clustering the spline representations
with k-means \citep{ABRAHAM2003}.
A similar 
approach makes use of the eigenfunctions of the covariance 
operator instead of B-splines \citep{PENGMULLER2008}.
In contrast, we will attempt to cluster functions or 
operators directly via a concentration of measure
approach similar to the previously described classification
procedure.

Consider 
the same setting to the previous section 
of multiple observations from multiple 
categories.  However, now the category labels are missing.
This is a functional mixture model where each observed functional 
datum is a stochastic process with one of $k$ possible covariance
operators.  In the below experiments, the data will be simulated
from a Gaussian process. 
The goal is to correctly separate the data into 
$k$ sets.  To achieve this, an expectation-maximization style
algorithm is implemented.

Let the observed operator data be $S_1,\ldots,S_n\in Op(L^2)$ 
where each $S_i = \text{cov}(f_1^{(i)},\ldots,f_{m_i}^{(i)})$ is
a rank $m_i$ operator produced from $m_i$ functional observations.
Let the latent label variables be $Y_1,\ldots,Y_n\in\{1,\ldots,k\}$.
Assuming no prior knowledge on the proportions of data in
each category, the algorithm is initialized with the 
Jeffreys prior for the Dirichlet distribution
by randomly generating
$
  \rho_{i,\cdot}^{(0)} \dist \distDirich{1/2,\ldots,1/2},
$  
the initial probability vector that $\Prob{Y_i = *}{f_i}$.

Assuming $t$ iterations of the algorithm have completed, 
we have a label probability vector $\rho_{i,\cdot}^{(t)}$ for 
each of the $n$ observations.
Given this collection of vectors, the expected proportions of each
category can be estimated as
$
  \tau_j^{(t+1)} = {n^{-1}}\sum_{i=1}^n \xval{\indc{Y_i=j}} =
  {n^{-1}}\sum_{i=1}^n \rho_{i,j}^{(t)}.
$
Similarly, a weighted sum of the data, $\hat{\Sigma}_j^{(t+1)}$,
and a weighted Rademacher sum, $R_j^{(t+1)}$,  can be used to update
the estimated covariance operators for each label $j$:
\begin{align*}
  \hat{\Sigma}_j^{(t+1)} &= 
  \frac{\sum_{i=1}^n\rho_{i,j}^{(t)}{S_i}}
  {\sum_{i=1}^n\rho_{i,j}^{(t)}},
  &
  R_j^{(t+1)} &= 
  \frac{ \sum_{i=1}^n \rho_{i,j}^{(t)}\veps_i
      \left({S_i}-\hat{\Sigma}_j^{(t+1)}\right) 
  }{\sum_{i=1}^n\rho_{i,j}^{(t)}}.
\end{align*}

Lastly,
a pooled weak variance is required, which is used in
place of each individual category weak variance.
Otherwise, in practice, one single category captures
all of the data points.  By defining the pooled 
covariance operator as 
$
  \hat{\Sigma}_\text{pool}^{(t+1)} =
  \sum_{j=1}^k \tau_j^{(t+1)} \hat{\Sigma}_j^{(t+1)},
$
then the pooled weak variance in the Gaussian case, for example,
is estimated by $2\lVert{\hat{\Sigma}_\text{pool}^{(t+1)}}\rVert_p$.

As a result, the label probability vectors $\rho_{i,\cdot}^{(t)}$ 
can be updated
given the $t+1$st collection of estimated covariance
operators, Rademacher sums, and the pooled covariance
operator.  From the previous section,
Equation~\ref{eqn:labelProb} can be used to determine 
$
  \rho_{i,j}^{(t+1)} =  \Prob{Y_i = j}
  {S_i,\hat{\Sigma}_1^{(t+1)},\ldots,\hat{\Sigma}_k^{(t+1)}},
$
the probability that observation $i$ belongs to the $j$th
category.  This process can be iterated until a local 
optimum is reached.

\section{Numerical Experiments}
\label{sec:numerics}
\subsection{Simulated and phoneme data}
\label{sec:numData}

To test each of the above three applications, experiments were
first run on simulated data.  These data sets were generated 
as observations from Gaussian or t-distributed processes.  
The covaraiance operators were randomly chosen given a specific
decay rate for the eigenvalues.

Secondly,
the phoneme data to be tested  \citep{FERRATYVIEU2003,HASTIE1995PDA} 
is a collection of 
400 log-periodograms for each of five different phonemes:
\phnaa as in the vowel of {``dark''}; 
\phnao as in the first vowel of {``water''}; 
\phndc as in the plosive of {``dark''}; 
\phniy as in the vowel of {``she''}; 
\phnsh as in the fricative of {``she''}.
Each curve contains the first 150 frequencies from a 32 ms
sound clip sampled at a rate of 16-kHz. 

\subsection{k Sample Comparison}
\label{sec:twoSampApp}

The above confidence set in Equation~\ref{eqn:nsampRegion} 
comparing $k$ samples can be used
to refute the null hypothesis that all covariance operators
are equal.  A two sample permutation test was performed in
\cite{PIGOLI2014}.  
Given two samples of functional data, $f_1^{(1)},\ldots,f_n^{(1)}$
and $f_1^{(2)},\ldots,f_m^{(2)}$ with associated covariance
operators $\Sigma^{(1)}$ and $\Sigma^{(2)}$, respectively,
the desired hypotheses to test are 
\begin{align*}
  \text{H}_0:& ~~\Sigma^{(1)}=\Sigma^{(2)}&
  \text{H}_1:& ~~\Sigma^{(1)}\ne\Sigma^{(2)}.
\end{align*}
When using a permutation test, the labels are randomly reassigned 
$M$ times, and each time, the distance between the two new 
covariance operators is computed.  For sufficiently large 
$M$, this procedure will return the exact significance
level of the observations with respect to the data set.

A power analysis was performed between the permutation method and
our proposed concentration approach using Equation~\ref{eqn:nsampRegion}.
Given two different operators $\Sigma^{(1)}$ and $\Sigma^{(2)}$
and $\gamma\ge0$,
an interpolation between the two operators is constructed as
$
  \Sigma^{(\gamma)} = [
    (\Sigma^{(1)})^{1/2} + 
    \gamma\{S(\Sigma^{(2)})^{1/2}-(\Sigma^{(1)})^{1/2}\}
  ][
    (\Sigma^{(1)})^{1/2} + 
    \gamma\{S(\Sigma^{(2)})^{1/2}-(\Sigma^{(1)})^{1/2}\}
  ]^{*},
$
where $S$ is an operator minimizing the Procrustes distance, 
between $\Sigma^{(1)}$ and~$\Sigma^{(2)}$, which is
{
  $ d_\text{Proc}(\Sigma^{(1)},\Sigma^{(2)})^2 = 
  \inf_{S\in U\{L^2(I)\}}\lVert{R^{(1)}-R^{(2)}S}\rVert_2^2$ where 
  $\Sigma^{(i)}=(R^{(i)})(R^{(i)})^*$ and 
  $U\left\{L^2(I)\right\}$ is the space of unitary operators on $L^2(I)$
  \citep{PIGOLI2014}.
} 

Monte Carlo simulations were run in order to estimate the power of
each test.  Two operators $\Sigma^{(1)}$ and $\Sigma^{(2)}$ with
similar eigenvalue decay were compared.
For sample size $n=50$, $\gamma \in \{0,.1,.2,.3,.4,.5\}$, and 
for sample size $n=500$, $\gamma\in\{0,.05,.1,.15,.2,.25\}$.
For each $\gamma$, 5000 samples of size $n$ were generated for 
$\Sigma^{(1)}$ and $\Sigma^{(\gamma)}$.  Equation~\ref{eqn:nsampRegion}
and the permutation method \citep{PIGOLI2014} were both implemented
to estimate the empirical power.

\begin{figure}
  \begin{center}
  \includegraphics[width=1\textwidth]{\PICDIR/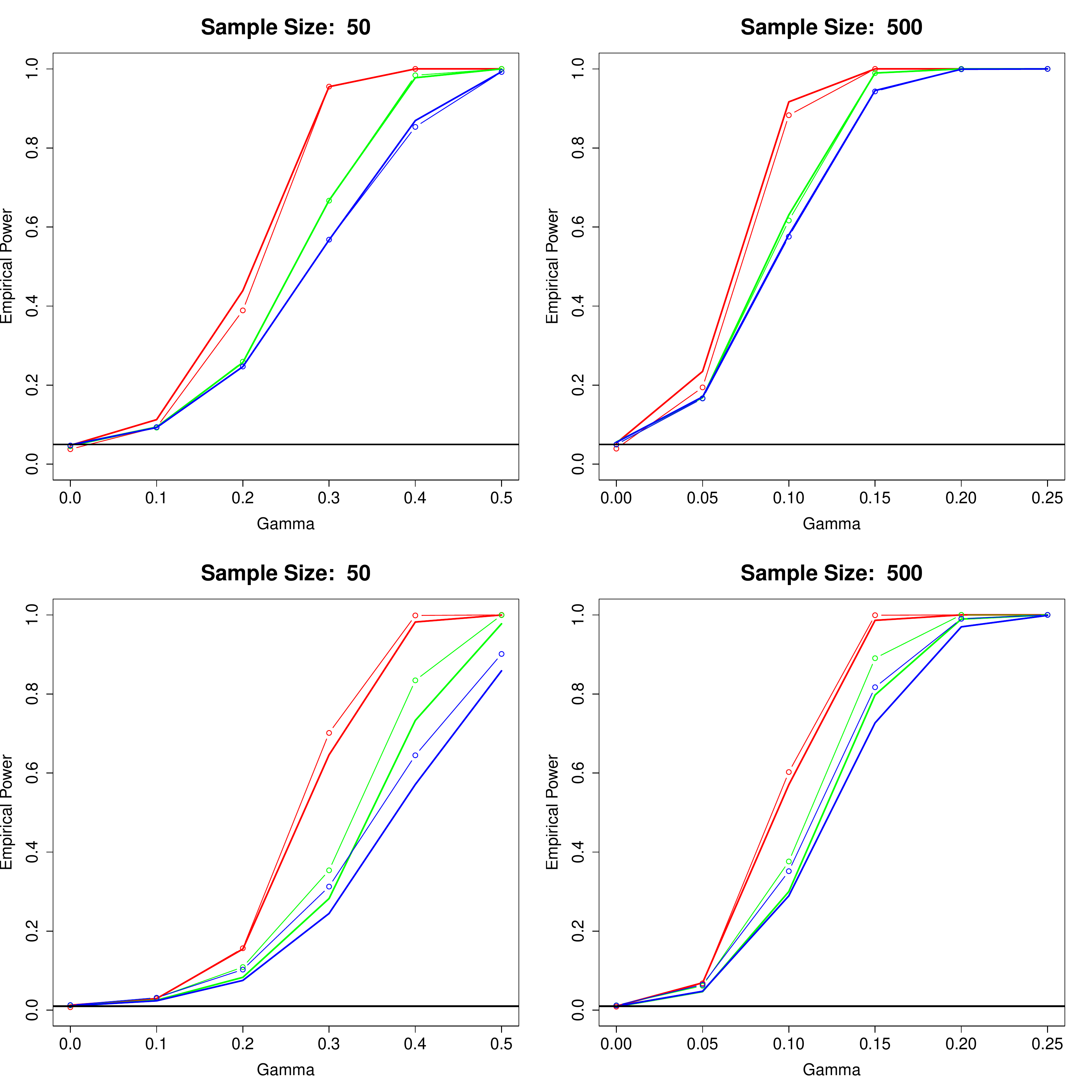}
  \end{center}
  \capt{
    \label{fig:powAn4}
    A power analysis for testing whether or not operator
    $\Sigma^{(1)} = \Sigma^{(\gamma)}$ comparing the
    permutation method (solid lines) with the concentration 
    approach (circle lines).
    The size $\alpha=0.05$ in 
    the top row, and $\alpha=0.01$ in the bottom.  
    The eigenvalues
    of the operators decay at a rate $O(k^{-4})$.  
    The red, green, and blue lines respectively correspond to 
    the trace class, Hilbert-Schmidt, and operator norms.
  }
\end{figure}

\begin{figure}
  \begin{center}
  \includegraphics[width=1\textwidth]{\PICDIR/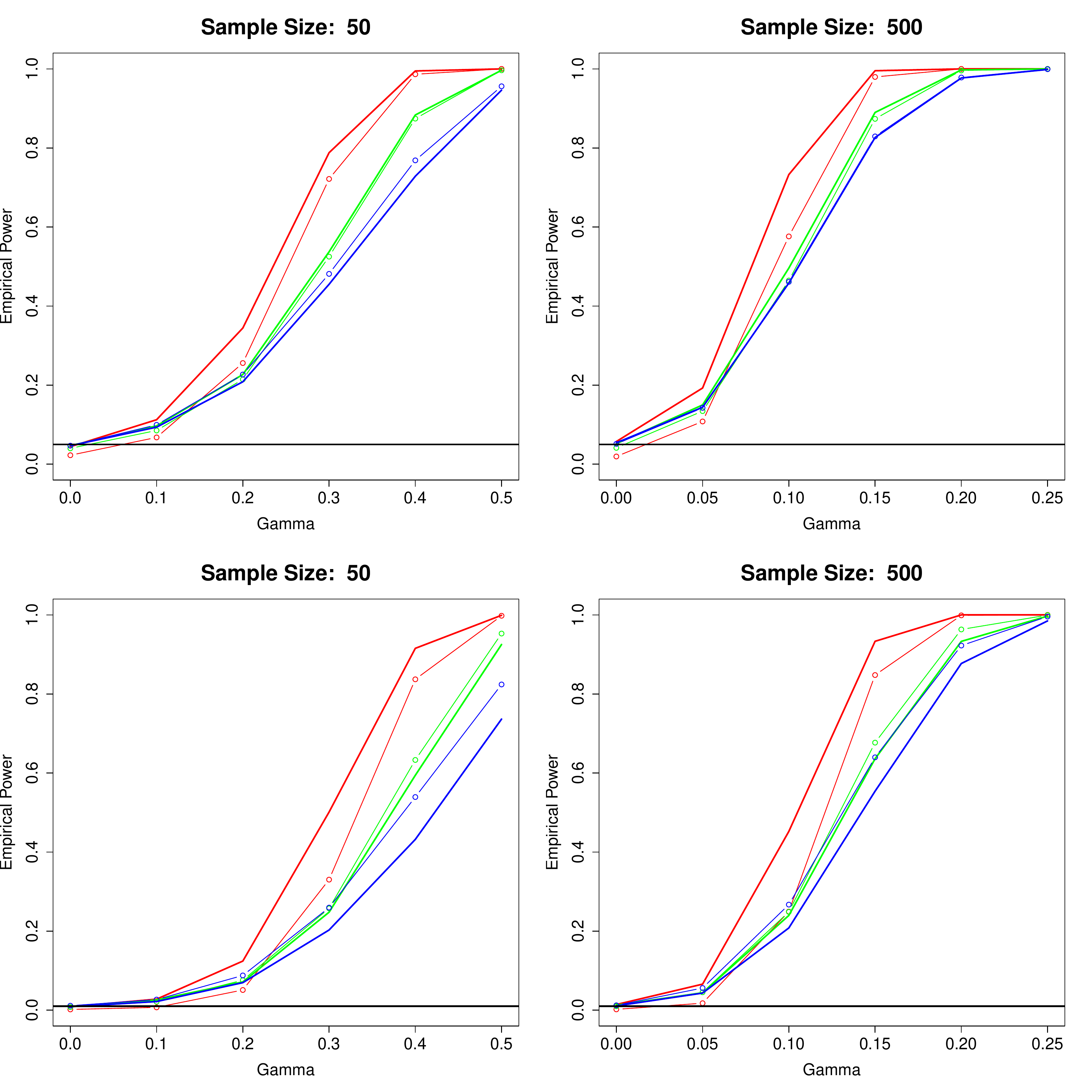}
  \end{center}
  \capt{
    \label{fig:powAn2}
    A power analysis for testing whether or not operator
    $\Sigma^{(1)} = \Sigma^{(\gamma)}$ comparing the
    permutation method (solid lines) with the concentration 
    approach (circle lines).
    The size $\alpha=0.05$ in 
    the top row, and $\alpha=0.01$ in the bottom.  
    The eigenvalues
    of the operators decay at a rate $O(k^{-2})$.  
    The red, green, and blue lines respectively correspond to 
    the trace class, Hilbert-Schmidt, and operator norms.
  }
\end{figure}

Figures~\ref{fig:powAn4} and~\ref{fig:powAn2} display the results
for operators whose eigenvalues decay at a quartic and quadratic rate,
respectively.  The solid lines indicate the power of the permutation
test, and the circle lines indicate the power of our concentration
approach.  The colors red, green, and blue correspond to the three
norms trace, Hilbert-Schmidt, and operator, respectively.

In most cases, the concentration approach is able to achieve 
the same power 
to reject the null as does the permutation test.  The notable
exception is for the trace norm when the eigenvalues decay slowly.
The added benefit to the concentration approach is the speed with 
which it executes.  Across all of the Monte Carlo simulations,
our concentration approach ran on average 28.14 times faster than 
the permutation method.  This was computed by tracking the amount
of computation time each method spent while producing the plots in 
Figures~\ref{fig:powAn4} and~\ref{fig:powAn2}, which corresponds
to $5$ values of $\gamma$, $2$ values of $\alpha$, 
$3$ different norms, $2$ sample sizes, and $5000$ replications each
resulting in 300,000 function calls for both the permutation and 
concentration methods.  Unlike the other norms, the 
Hilbert-Schmidt norm can be calculated without explicit
computation of the eigenvalues.
For each evaluation of the permutation test, 100 permutations of the
data were generated, which corresponds to 100 random draws and 
100 eigenvalue computations.
More accuracy would require even more permutations.
In comparison, our concentration approach requires only $3k$ eigenvalue
computations and no random draws and hence is only dependent on the 
number of samples regardless of data size or $\alpha$.

The proposed $k$-sample test was also used to compare 
samples of log-periodogram curves from the spoken 
phonemes \phnaa and 
\phnao.  As one
can imagine, these vowels can be hard to distinguish;
see Section~\ref{sec:emAlgPractice} for further evidence of this.
For $k \in \{2,3,4,5,6\}$, $k-1$ disjoint sets of 
40 \phnaa curves and one set of 40 \phnao curves 
were randomly sampled from the data set.  This was
replicated 500 times, and each time
Equation~\ref{eqn:nsampRegion} was used to decide
whether or not the $k$ covariance operators were equivalent at the 
$\alpha=0.05$ level.
The resulting estimated statistical power for each $k$ is
\begin{center}
\begin{tabular}{cccccc}
  $k$   &  2 & 3 & 4 & 5 & 6 \\
  Power &  0.00 & 0.018 & 0.228 & 0.656 & 0.936
\end{tabular}
\end{center}
In the null setting, the above experiement was rerun except
that every disjoint set of curves came from the \phnaa set.
The resulting experimentally computed test sizes are
\begin{center}
\begin{tabular}{cccccc}
  $k$   &  2 & 3 & 4 & 5 & 6 \\
  Size  &  0.00 & 0.00 & 0.00 & 0.004 & 0.072
\end{tabular}
\end{center}

\subsection{Binary and trinary classification}
\label{sec:simData}

Our concentration of measure (CoM) method is implemented 
on covariance operators making use of the trace norm
$\norm{\cdot}_\text{tr}$ where for a covariance operator 
$\Sigma$ with eigenvalues $\{\lmb_i\}_{i=1}^\infty$, 
$\norm{\Sigma}_\text{tr} = \sum_{i=1}^n \abs{\lmb_i}$.
The trace norm was chosen based on the analysis of 
the preceding section as well as that 
of \cite{PIGOLI2014}
where it achieved the best performance when compared with
the other p-Schatten norms.
The CoM approach to classification of operators
is tested in a variety of simulations against other
standard approaches to functional classification.
The methods used for comparison are $k$-nearest 
neighbours \citep{FERRATYVIEU2006}, classification 
using kernel estimators \citep{FERRATYVIEU2003},
general linear model \citep{MULLERSTADTMULLER2005}, 
and regression trees.

The first simulation asks each method to classify observed
mean zero Gaussian process data or mean zero t-process data 
with 4 degrees of freedom.
The two covariance operators in question, 
$\Sigma_1$ and $\Sigma_2$, are  
the sample covariances of the male and of the females of the 
Berkeley growth curve data \citep{RAMSAYSILVERMAN2005}.
In particular, $n$ collections of $k$ curves were generated
from each of $\Sigma_1$ and $\Sigma_2$ as a training set,
and $m$ collections of $k$ curves were generated as a test
set.
The CoM method was trained on the set of $n$ sample 
covariances and used to classify
each of the $m$ test covariances.  
The remaining classification methods were trained and tested in two
separate ways:  By treating each sample covariance as a function
and classifying as usual,
and by training on all $n\times k$ observations and testing
each of the $m$ collections by classifying each constituent
curve individually and taking a majority vote with
ties settled by a uniform random draw.  

For group sizes $k=1,2,4,8,16$, 100 simulations were run
with $n=100$ sets of $k$ training curves.  
To compare the accuracy of each approach  
$m=100$ sets of $k$ testing curves were generated for each operator.  
The accuracy of each method
is tabulated in Table~\ref{tab:classBinary}.

The concentration method performed well against the alternatives.
Its performance was on par with the kernel method 
applied to each covariance operator as a function.
Our method was only consistently outperformed by the 
kernel method implementing the majority vote approach.  However,
the two operators in question have very similar weak 
variances.  The next simulation demonstrates how the 
concentration method adapts naturally when the variances
of each label significantly differ. 

\begin{table}
  \caption{
    \label{tab:classBinary} 
    A comparison of the performances of five classification 
    methods.
    The first entry for each 
    method corresponds to classifying the covariance operators 
    as functions.  The prime entry corresponds to classifying curves
    with a majority vote.
    The estimated percent of correct classification 
    is listed in the table with the sample standard deviation in 
    brackets.  The top block comes from Gaussian process data, and
    the bottom comes from t-process data with 4 degrees of freedom.
    The highest percentage of each column is marked in bold.
  }
  \centering
  \begin{tabular}{llllll}
    $k$ & \multicolumn{1}{c}{1} & \multicolumn{1}{c}{2} & 
           \multicolumn{1}{c}{4} & \multicolumn{1}{c}{8} & 
           \multicolumn{1}{c}{16} \\
    CoM &     0.62 (0.05) & 0.62 (0.05) &
      0.76 (0.08) & 0.87 (0.06) & 0.96 (0.03) \\
    KNN &     0.52 (0.04) & 0.44 (0.04) & 
      0.57 (0.04) & 0.76 (0.04) & 0.91 (0.02) \\
    KNN$'$&  \multicolumn{1}{c}{$\cdot$}   & 0.47 (0.03) &  
      0.59 (0.04) & 0.74 (0.03) & 0.89 (0.02)\\
    Kernel &  {\bf0.65} (0.04) & 0.64 (0.05) & 
      0.75 (0.03) & 0.87 (0.03) & 0.96 (0.02) \\
    Kernel$'$& \multicolumn{1}{c}{$\cdot$} &{\bf0.70} (0.04) & 
      {\bf0.82} (0.02) & {\bf0.92} (0.02) & {\bf0.99} (0.01) \\
    GLM &     0.51 (0.04) & 0.62 (0.04) & 
      0.74 (0.04) & 0.86 (0.02) & 0.94 (0.02)\\
    GLM$'$&  \multicolumn{1}{c}{$\cdot$}   &0.50 (0.04)& 
      0.50 (0.03) & 0.50 (0.04) & 0.50 (0.04)\\
    Tree &    0.57 (0.04) & 0.54 (0.04) &
      0.59 (0.03) & 0.66 (0.04) & 0.75 (0.03)\\
    Tree$'$& \multicolumn{1}{c}{$\cdot$}   &0.55 (0.04)&
      0.59 (0.04) & 0.60 (0.05) & 0.60 (0.09)\\\\
    CoM &    0.59 (0.05)  & 0.62 (0.05) &
      0.75 (0.07) & 0.86 (0.06)  & 0.95 (0.04) \\
    KNN &    0.42 (0.04)  & 0.45 (0.04) & 
      0.58 (0.04) & 0.76 (0.03) & 0.92 (0.02) \\
    KNN$'$& \multicolumn{1}{c}{$\cdot$}  & 0.45 (0.04)  & 
      0.58 (0.04) & 0.72 (0.03) & 0.89 (0.03) \\
    Kernel & {\bf0.65} (0.04)  & 0.64 (0.05)  & 
      0.75 (0.04) & 0.87 (0.02) & 0.96 (0.02)  \\
    Kernel$'$& \multicolumn{1}{c}{$\cdot$} & {\bf0.68} (0.04) & 
      {\bf0.80} (0.03) & {\bf0.92} (0.02) & {\bf0.99} (0.01)  \\
    GLM &    0.50 (0.03)  & 0.62 (0.03) & 
      0.74 (0.04) & 0.86 (0.03) & 0.94 (0.02) \\
    GLM$'$& \multicolumn{1}{c}{$\cdot$}  & 0.50 (0.03) & 
      0.50 (0.03) & 0.51 (0.04) & 0.50 (0.03) \\
    Tree &   0.54 (0.04)  & 0.54 (0.04) &
      0.59 (0.03) & 0.67 (0.03) & 0.75 (0.03) \\
    Tree$'$& \multicolumn{1}{c}{$\cdot$} & 0.54 (0.04) &
      0.57 (0.04) & 0.59 (0.05) & 0.57 (0.06)
  \end{tabular}
  \vspace{0.1in}

  CoM, concentration of measure; KNN, k-nearest-neighbours;
  GLM, generalized linear model; Tree, regression tree.
\end{table}


Continuing from the previous simulation, a third operator is
constructed from $\Sigma_1$ and $\Sigma_2$ by averaging these
two and then scaling up the non-principle eigenvalues by
a factor of 5.  This, in some sense, creates a third operator
between the first two, but with higher variance.  The simulation
is carried out precisely as before, but incorporating all three
operators.  In this setting, our concentration approach demonstrates
the best performance.  The results are listed in 
Table~\ref{tab:classTrinary}.

\begin{table}
  \caption{
    \label{tab:classTrinary} 
    A comparison of the performances of five classification 
    methods as in Table~\ref{tab:classBinary}, but with three
    potential classes from which to choose.
    The estimated percent of correct classification 
    is listed in the table with the sample standard deviation in 
    brackets.  The top block comes from Gaussian process data, and
    the bottom comes from t-process data with 4 degrees of freedom.
    The highest percentage of each column is marked in bold.
  }
  \centering
  \begin{tabular}{llllll}
    $k$ & \multicolumn{1}{c}{1} & \multicolumn{1}{c}{2} & 
           \multicolumn{1}{c}{4} & \multicolumn{1}{c}{8} & 
           \multicolumn{1}{c}{16} \\
    CoM &  0.51 (0.04) & 0.55 (0.04) & {\bf0.75} (0.05)
       & {\bf0.89} (0.05)  & {\bf0.97} (0.03)  \\
    KNN &  0.50 (0.03) & 0.52 (0.03) & 0.61 (0.03)
       & 0.75 (0.03)  & 0.90 (0.02) \\
    KNN$'$&  \multicolumn{1}{c}{$\cdot$}  & 0.55 (0.03)  & 0.68 (0.03) 
       & 0.80 (0.02) & 0.90 (0.02) \\
    Kernel & {\bf0.54} (0.03) & 0.52 (0.03) & 0.64 (0.03)
       & 0.77 (0.03)  & 0.92 (0.02) \\
    Kernel$'$& \multicolumn{1}{c}{$\cdot$} & {\bf0.58} (0.03) & 0.69 (0.02)
       & 0.81 (0.03) & 0.92 (0.02) \\
    GLM &  0.36 (0.04) & 0.41 (0.04) & 0.49 (0.04)
       & 0.57 (0.03)  & 0.65 (0.03) \\
    GLM$'$&  \multicolumn{1}{c}{$\cdot$}  & 0.35 (0.04) & 0.36 (0.04)
       & 0.36 (0.05) & 0.35 (0.05) \\
    Tree & 0.44 (0.03) & 0.44 (0.03) & 0.45 (0.03)
       & 0.50 (0.03)  & 0.55 (0.03) \\
    Tree$'$& \multicolumn{1}{c}{$\cdot$} & 0.46 (0.03) & 0.51 (0.04)
       & 0.51 (0.07) & 0.47 (0.07) \\\\
    CoM &  0.46 (0.05)    & 0.50 (0.06)  & 0.63 (0.05)
       & 0.75 (0.08)  & 0.85 (0.07) \\
    KNN &  0.46 (0.03)    & 0.49 (0.03)  & 0.57 (0.03)
       & 0.67 (0.03)  & 0.78 (0.02) \\
    KNN$'$& \multicolumn{1}{c}{$\cdot$}  & 0.50 (0.03)  & 0.64 (0.03)
       & {\bf0.77} (0.02)  & {\bf0.87} (0.02) \\
    Kernel & {\bf0.50} (0.03)  & 0.48 (0.03)  & 0.57 (0.03)
       & 0.68 (0.03)  &  0.80 (0.02) \\
    Kernel$'$& \multicolumn{1}{c}{$\cdot$} & {\bf0.53} (0.03)  & {\bf0.66} (0.03)
       & {\bf0.77} (0.02)  &  0.85 (0.02) \\
    GLM &  0.35 (0.03)    & 0.41 (0.04)  & 0.46 (0.04)
       & 0.53 (0.03)  & 0.58 (0.04) \\
    GLM$'$& \multicolumn{1}{c}{$\cdot$}  & 0.35 (0.03)  & 0.36 (0.04)
       & 0.36 (0.04)  & 0.36 (0.06) \\
    Tree &  0.42 (0.03)   & 0.44 (0.03)  & 0.45 (0.03)
       & 0.46 (0.03)  & 0.49 (0.03) \\
    Tree$'$& \multicolumn{1}{c}{$\cdot$} & 0.43 (0.03)  & 0.47 (0.04)
       & 0.48 (0.06) & 0.46 (0.08)
  \end{tabular}
  \vspace{0.1in}

  CoM, concentration of measure; KNN, k-nearest-neighbours;
  GLM, generalized linear model; Tree, regression tree.
\end{table}

These five methods tested on simulated data 
were also tested 
against phoneme data.  
Across 50 iterations, each set of 400 curves was partitioned at random
into an 100 curve training set and a 300 curve testing set.
The five classifiers were trained and run on each of the 
$300\times5$ curves individually.  For our concentration of measure
approach, the rank one operator associated to each individual curve
was compared with the covariance operator formed from the $100\times5$
training curves.
The results are detailed in
Table~\ref{tab:phonemeClass}.  Our concentration of measure 
approach only uniformly outperforms the regression tree classifier,
but has comparable performance to the other three methods, and none
of the competing methods uniformly outperforms ours.

\begin{table}
  \caption{
    \label{tab:phonemeClass}
    Percentage of correct classification of the five
    phonemes against the five methods.
    The highest percentage of each column is marked in bold.
  }
  \centering
  \begin{tabular}{llllll}
    & \multicolumn{1}{c}{\phnaa} & \multicolumn{1}{c}{\phnao}
    & \multicolumn{1}{c}{\phndc}& \multicolumn{1}{c}{\phniy}
    & \multicolumn{1}{c}{\phnsh} \\
    CoM    & 0.769 & 0.768 & 0.966 & {\bf0.985} & 0.994 \\
    KNN    & 0.724 & 0.791 & {\bf0.985} & 0.974 & {\bf1.000} \\
    Kernel & 0.720 & {\bf0.805} & 0.984 & 0.972 & 0.999 \\
    GLM    & {\bf0.790} & 0.723 & 0.982 & 0.959 & 0.992 \\
    Tree   & 0.708 & 0.694 & 0.956 & 0.878 & 0.926 
  \end{tabular}
\end{table}

\subsection{The expectation-maximization algorithm in practice }
\label{sec:emAlgPractice}

The experiments described and depicted below 
make use of the trace norm only.
It was determined through experimentation 
that the expectation-maximization algorithm we propose
in Section~\ref{sec:emalg} does 
not perform well under the topology
of either the Hilbert-Schmidt or operator norms
as they give more emphasis to the principle eigenvalue
at the expense of the others.
The usual behavior under these norms is for all estimates
to converge to the average of all of the data points.
This is in contrast to the better performance of the 
algorithm making use of the
trace norm, which is somewhat more
uniform in its treatment of the eigenstructure.

As a first test case, this algorithm was run given three target 
covariance operators, which were constructed by taking three
randomly generated orthonormal bases $U_i$ and a diagonal operator $D$ 
of eigenvalues decaying at a rate
$\lmb_k=O(k^{-4})$ and multiplying $\Sigma_i = U_iD\TT{U_i}$.  
Let the three target covariance operators be
denoted as $\Sigma_a$, $\Sigma_b$, and $\Sigma_c$.  For each
of these operators, 1000 rank four data points were generated 
from a zero mean Gaussian process.  
From the data, the algorithm initializes
three estimates $\hat{\Sigma}_1^{(t)}$, $\hat{\Sigma}_2^{(t)}$, and
$\hat{\Sigma}_3^{(t)}$, which attempt to locate the three target 
operators as the method iterates.
After 15 iterations,
the original 3000 data points were perfectly separated into
three groups. 
To make the problem harder, 
a second test case was run identical to the first except that the
observed operators are all of rank one.  
The resulting clusters from both tests are in 
Table~\ref{tab:clustTest1}.
The inaccuracy in the rank one setting is equivalent
to the poor performance of classification of rank one operators 
detailed in Tables~\ref{tab:classBinary} and~\ref{tab:classTrinary}.

\begin{table}
\caption{
  \label{tab:clustTest1}
  Clustering of simulated operators.
}
\centering
\begin{tabular}{lccccccc}
  & \multicolumn{3}{c}{Rank 4 Operators} &
  & \multicolumn{3}{c}{Rank 1 Operators} \\
  & Cluster 1 & Cluster 2 & Cluster 3 &
  & Cluster 1 & Cluster 2 & Cluster 3\\
  Label $a$ & 0 & 0 & 1000 & & 169 & 557 & 274\\
  Label $b$ & 0 & 1000 & 0 & & 184 & 549 & 267\\
  Label $c$ & 1000 & 0 & 0 & & 483 & 230 & 287
\end{tabular}
\end{table}

For the phoneme data 
all 400 sample curves from each of the 
five phoneme sets were clustered individually as curves.  
The algorithm was run for 20 iterations
and told to partition the data into five clusters. 
The results are in Table~\ref{tab:clustPhon1}.
Cluster~3 captured almost all of the 
vowels \phnaa and \phnao, which, recalling their definition in
Section~\ref{sec:numData}, are 
quite similar in sound.  Clusters~2 and~4 contain the 
majority of \phnsh and \phndc curves, respectively.  
Lastly, Clusters~1 and~5 split the set of \phniy curves
roughly in half.
Rerunning the experiment with only four clustered nicely 
separated the data as displayed in Table~\ref{tab:clustPhon2}.
Cluster~3 again contains almost all of the \phnaa and \phnao.
The remaining clusters mostly partition the \phndc, \phniy, 
and \phnsh with high accuracy.  
Hence, barring the ability to distinguish the very 
similar \phnaa and \phnao curves, the 
proposed expectation-maximization algorithm is an effective 
method for the unsupervised clustering of phonemes.

\begin{table}
\caption{
  \label{tab:clustPhon1}
  Clustering 2000 phoneme curves into 5 clusters.
}
\centering
\begin{tabular}{lccccc}
  & Cluster 1 & Cluster 2  & Cluster 3 & Cluster 4 & Cluster 5  \\
  \phnaa &   3 &   0 & 397 &   0 &   0   \\
  \phnao &   0 &   0 & 397 &   1 &   2   \\
  \phndc &   0 &   0 &   0 & 378 &  22   \\
  \phniy & 212 &   0 &   0 &   1 & 187   \\
  \phnsh &   4 & 393 &   0 &   0 &   3   
\end{tabular}
\end{table}
\begin{table}
\caption{
  \label{tab:clustPhon2}
  Clustering 2000 phoneme curves into four clusters.
}
\centering
\begin{tabular}{lcccc}
  & Cluster 1 & Cluster 2  & Cluster 3 & Cluster 4 \\
  \phnaa &   0&   0& 400&   0   \\
  \phnao &   1&   0& 398&   1   \\
  \phndc & 384&   1&   0&  15   \\
  \phniy &   1&   5&   0& 394   \\
  \phnsh &   0& 397&   0&   3   
\end{tabular}
\end{table}

\bibliographystyle{plainnat}
\bibliography{\BIBDIR/kasharticle,\BIBDIR/kashbook}

\end{document}